\documentclass[nohyperref]{article}

\usepackage{microtype}
\usepackage{graphicx}
\usepackage{subfigure}
\usepackage{booktabs} % for professional tables

\usepackage{hyperref}

\usepackage[accepted]{icml2022}
\usepackage[subtle]{savetrees}

\usepackage{amsmath}
\usepackage{amssymb}
\usepackage{mathtools}
\usepackage{amsthm}

\usepackage[capitalize,noabbrev]{cleveref}

\theoremstyle{plain}

\theoremstyle{definition}

\theoremstyle{remark}

\usepackage[textsize=tiny]{todonotes}

\usepackage{aas_macros}

\icmltitlerunning{A Convolutional Neural Network Approach to Supernova Time-Series Classification}

\begin{document}

\twocolumn[
\icmltitle{A Convolutional Neural Network Approach to Supernova Time-Series Classification}



\icmlsetsymbol{equal}{*}

\begin{icmlauthorlist}
\icmlauthor{Helen Qu}{penn}
\icmlauthor{Masao Sako}{penn}
\icmlauthor{Anais M{\"o}ller}{sch}
\icmlauthor{Cyrille Doux}{grenoble}
\end{icmlauthorlist}

\icmlaffiliation{penn}{Department of Physics and Astronomy, University of Pennsylvania, Philadelphia, PA 19104, USA}
\icmlaffiliation{grenoble}{Université Grenoble Alpes, CNRS, LPSC-IN2P3, 38000 Grenoble, France}
\icmlaffiliation{sch}{Centre for Astrophysics and Supercomputing, Swinburne University of Technology, 31122 Hawthorn, VIC, Australia}

\icmlcorrespondingauthor{Helen Qu}{helenqu@sas.upenn.edu}

\icmlkeywords{Supernova, Convolutional Neural Networks, Classification}

\vskip 0.3in
]



\printAffiliationsAndNotice{}  % leave blank if no need to mention equal contribution

\begin{abstract}
One of the brightest objects in the universe, supernovae (SNe) are powerful explosions marking the end of a star's lifetime. Supernova (SN) type is defined by spectroscopic emission lines, but obtaining spectroscopy is often logistically unfeasible. Thus, the ability to identify SNe by type using time-series image data alone is crucial, especially in light of the increasing breadth and depth of upcoming telescopes. We present a convolutional neural network method for fast supernova time-series classification, with observed brightness data smoothed in both the wavelength and time directions with Gaussian process regression. We apply this method to full duration and truncated SN time-series, to simulate retrospective as well as real-time classification performance. Retrospective classification is used to differentiate cosmologically useful Type Ia SNe from other SN types, and this method achieves $>99\%$ accuracy on this task. We are also able to differentiate between 6 SN types with 60\% accuracy given only two nights of data and 98\% accuracy retrospectively.
\end{abstract}

\section{Introduction}
\label{intro}

Supernovae are extremely bright explosions of stars that can rival the luminosity of an entire galaxy before fading away in just a few weeks. Current theories predict that supernovae are produced via two primary physical processes: gravitational collapse of the core of a massive star, and sudden re-ignition of nuclear fusion in a white dwarf. These different pathways produce SNe with different properties, which is encompassed in their type designation. Determining the type of observed SNe is particularly important in the case of type Ia SNe. Unlike other SN types, their standardizable brightness makes them excellent distance indicators (or \textit{standard candles}) on cosmological scales. Type Ia SNe were instrumental in the Nobel Prize-winning discovery of dark energy and the accelerating expansion of the universe \cite{perlmutter,riess} and continue to be a crucial tool for further understanding the nature of dark energy and the composition of our universe.

Modern-scale sky surveys, including SDSS, Pan-STARRS, and the Dark Energy Survey, have identified thousands of supernovae throughout their operational lifetimes \cite{sdss,panstarrs,des}. However, it has been logistically challenging to follow up on most of these detections spectroscopically. The result is a small number of SNe with spectroscopically confirmed types and a large dataset of SN candidates with only observed brightness data in broad wavelength bands (\textit{photometry}) and no type information. The upcoming Rubin Observatory Legacy Survey of Space and Time (LSST) is projected to discover $10^7$ supernovae \cite{lsstsciencecollaboration2009lsst}, with millions of transient alerts awaiting classification each observing night. A reliable photometric classification algorithm will allow us to tap into the vast potential of the photometric dataset and pave the way for confident classification and analysis of the ever-growing library of transients from current and future sky surveys.

Given the scarcity of spectroscopic resources, the goal of this work is twofold: (1) confident retrospective classification of type Ia SNe using photometry alone in order to increase the number of cosmologically useful SNe, and (2) reliable real-time determination of SN type early in its lifetime to aid in optimal allocation of spectroscopic resources. 

Our model produces both retrospective and real-time classification results with photometry data in the form of \textit{light curves}, time-series data of observed brightnesses in multiple wavelength bands. Other deep learning approaches to this task include recurrent neural network-based SuperNNova \cite{supernnova} and RAPID \cite{rapid}, CNN-based PELICAN \cite{pasquet}, and recurrent autoencoder-based SuperRAENN \cite{superraenn}.

\begin{figure*}[ht]
    \begin{center}
    \includegraphics[scale=0.2,trim={0cm 3cm 0cm 3cm},clip]{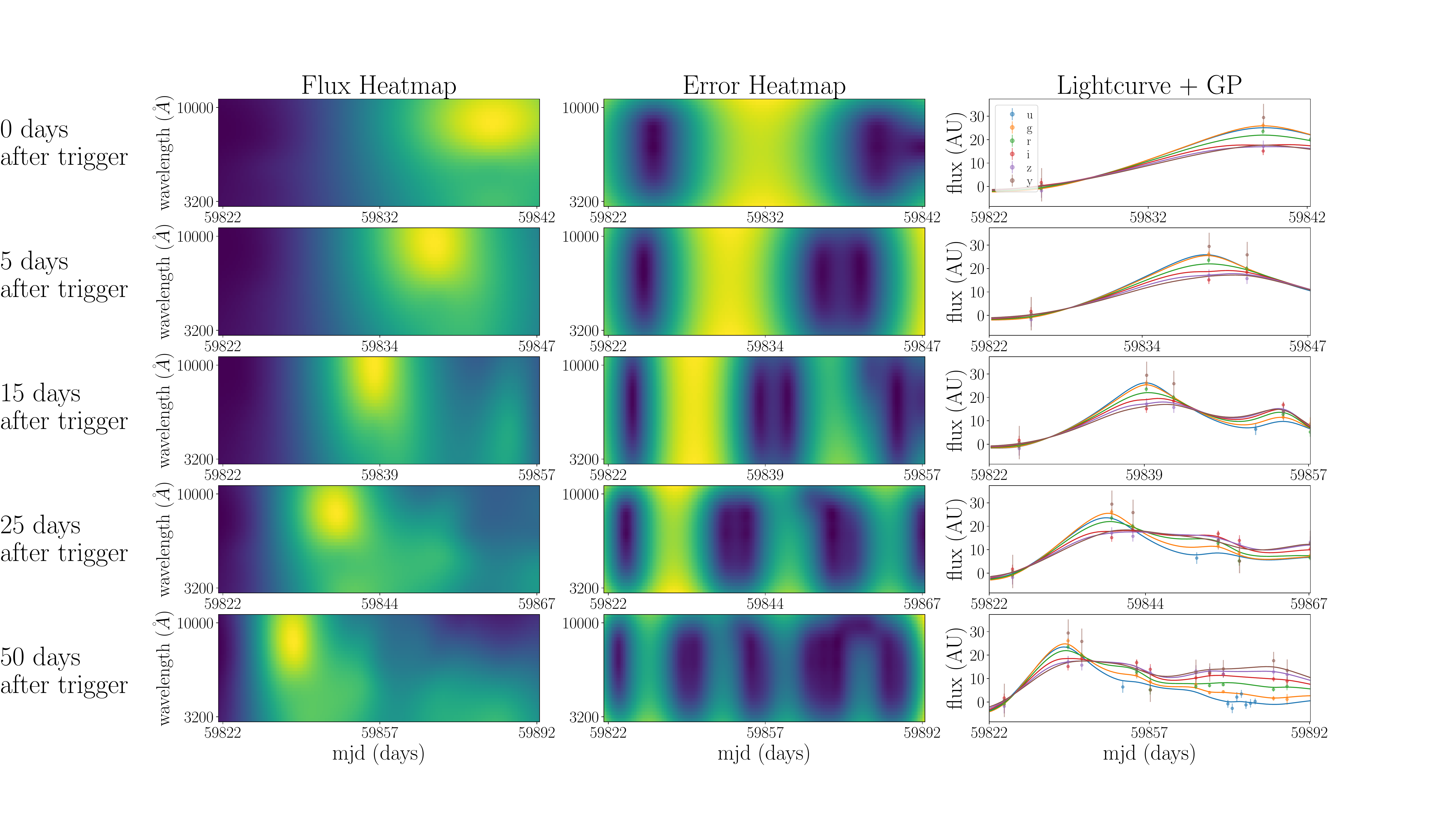}
    \centering
    \caption{Flux and error heatmaps as well as lightcurves and Gaussian process fits of an SNII ($z=0.39$) shown truncated at 0, 5, 15, 25, and 50 days after trigger. The flux and flux error measurements from the raw photometry are shown as points with error bars, while the Gaussian process fits to each photometry band are shown as curves.}
    \label{lightcurves}
    \end{center}
\end{figure*}

\section{Data}
\subsection{Datasets}
This model was trained and evaluated on datasets created for the 2018 Photometric LSST Astronomical Time Series Classification Challenge (PLAsTiCC) \cite{plasticc}. All results were produced with a 80\%/10\%/10\% training/validation/test split.

All lightcurves were observed in LSST's $ugrizY$ bands and realistic observing conditions were simulated using the LSST Operations Simulator \cite{delgado}. We selected all type II, Iax, Ibc, Ia-91bg, Ia, and SLSN-1 supernova sources from this dataset and chose only supernovae in the well-sampled deep drilling fields. For binary (Ia vs. non-Ia) classification, the total source count is 12,611 after basic quality cuts on the sources: each source must have at least 5 detections, have cumulative signal-to-noise ratio at least 10, and have observations spanning at least 30 days.

The dataset used for type classification was created using the SuperNova ANAlysis package (SNANA) \cite{snana} with the same parameters as the main dataset. 7,685 examples of each of the 6 types listed above were selected, making a full dataset size of 46,110 examples. To evaluate the model's real-time classification performance, five datasets were created from these simulations starting 20 nights prior to the date of trigger ($t_{\mathrm{trigger}}$) and end at $N=$ 0, 5, 15, 25, and 50 days after the date of trigger, respectively, where \textit{trigger} is defined as the next observation exceeding a $5\sigma$ signal-to-noise detection threshold that occurs at least one night after the first.

\begin{figure*}
    \includegraphics[scale=0.50, trim={2.25cm 0cm 0cm 0cm}]{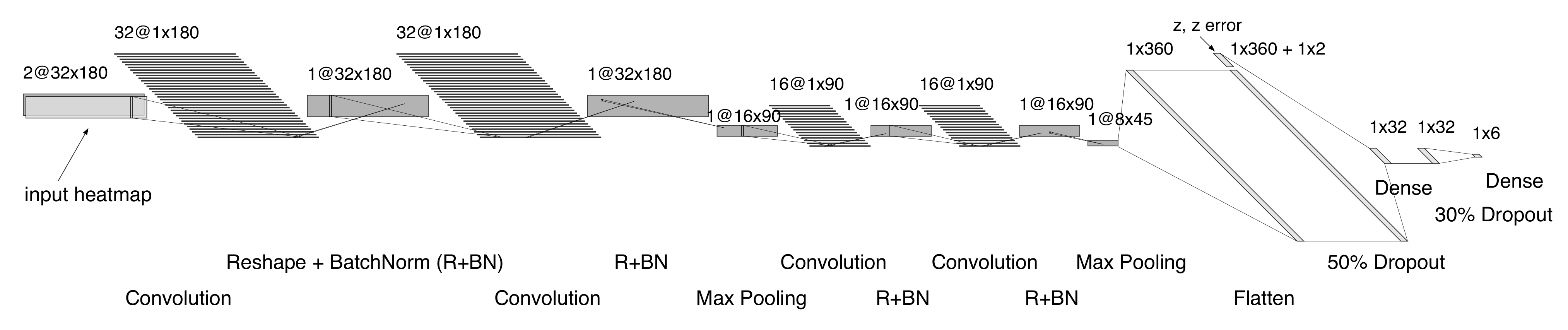}
    \centering
    \caption{Model architecture with redshift information for real-time typing. The base architecture for retrospective binary classification does not include the redshift input (labeled $z, z_{\mathrm{error}}$) and the output layer is replaced by a single sigmoid function.}
    \label{architecture}
\end{figure*}

\subsection{Preprocessing}
Prior to training, we preprocess our lightcurve data into 2-dimensional ``images". First, all observations are labeled with the filter's central wavelength, calculated from the LSST filter transmission functions. We then follow Avocado \cite{Boone_2019} to apply 2-dimensional Gaussian process regression to the raw lightcurve data to model the event in the wavelength ($\lambda$) and time ($t$) dimensions. We obtain predictions from the trained Gaussian process model on a grid of $\lambda,t$ values and call this our ``image". The input to the model is the grid of Gaussian process predictions stacked depthwise with the Gaussian process prediction uncertainties at each $\lambda,t$, all normalized to the maximum flux value.

\cref{lightcurves} shows the raw lightcurve data, the Gaussian process model, and the resulting flux and uncertainty heatmaps for each of the real-time classification datasets.

\section{Model}
\subsection{Base Model Architecture}

The relatively simple architecture of our model, shown in \cref{architecture}, allows for a minimal number of trainable parameters, speeding up the training process significantly without compromising on performance. It has a total of $\sim$ 22,000 trainable parameters when trained on images of size $32 \times 180 \times 2$ $(h \times w \times d)$.

All convolutional layers in our model use full-height kernels, i.e. each convolutional layer has $h$ filters and a kernel size of $h \times 3$. This choice allows the model to learn from observations in all wavelengths simultaneously at every point in time. We reshape the resulting feature map to $h \times d \times 1$ to allow for further convolutions. The output layer is a single node for binary classification and 6 nodes for typing. \cref{architecture} shows the version for typing.

For all reported results, the model is trained with crossentropy loss and optimized with Adam \cite{kingma2017adam} at a constant 1e-3 learning rate for 400 epochs.

\subsection{Model Architecture with Redshift}
Cosmological redshift describes how quickly a light source is moving away from the observer, similar to the Doppler effect. Redshift is correlated with brightness and can help break degeneracies between types, but is not always available. To avoid relying on precise redshift estimates, all retrospective binary classification results are reported without redshift information. Real-time typing results are reported with and without redshift information.

\begin{figure}
    \includegraphics[scale=0.35]{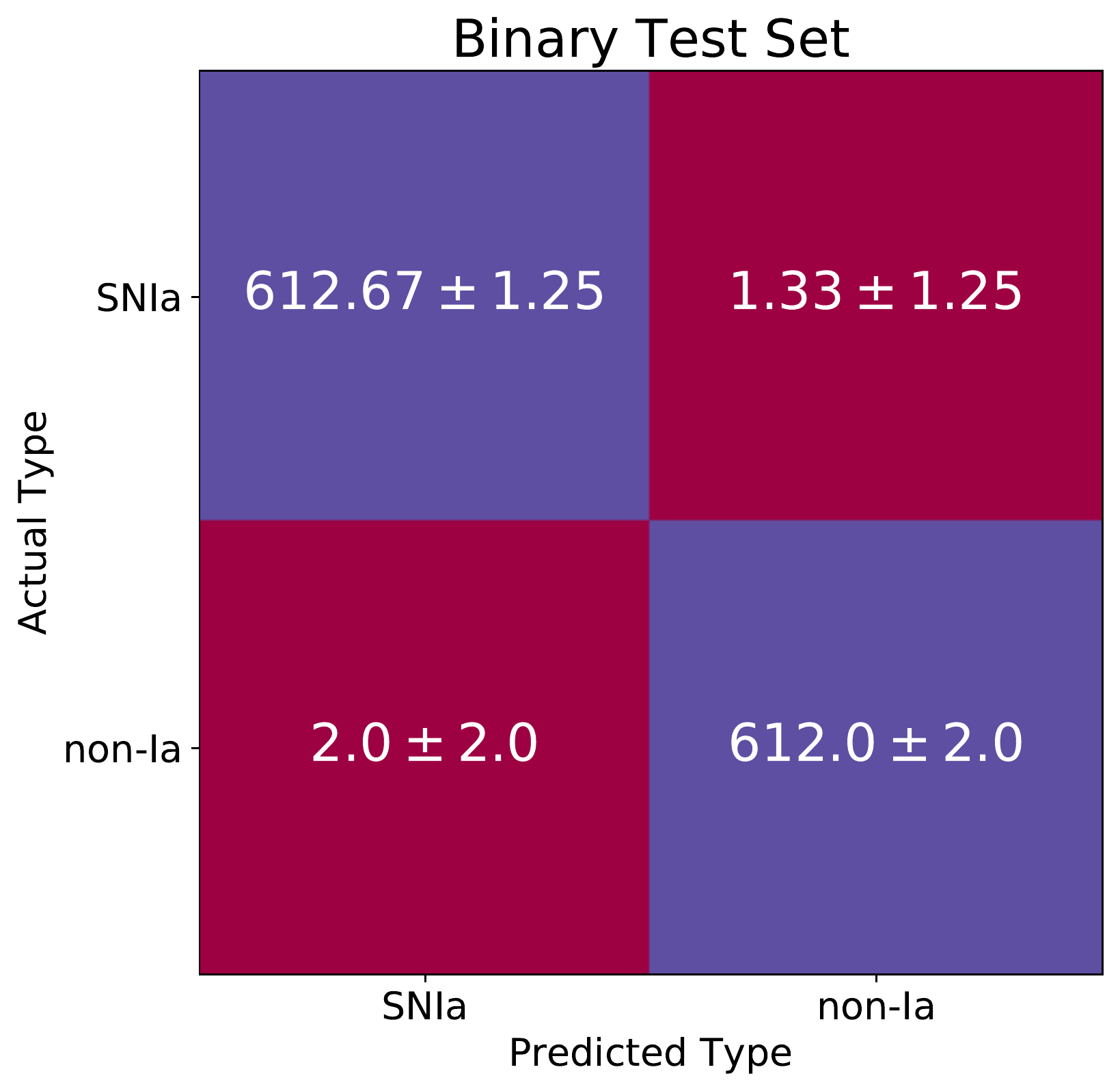}
    \centering
    \caption{Confusion matrix showing average and standard deviation over five runs for binary classification on the test set. The absolute numbers of examples are shown here as opposed to percentages.}
    \label{binary-confusion}
\end{figure}
%TODO: architecture with/without redshift together into one fig
The architecture of our model incorporating redshift and redshift uncertainty is shown in \cref{architecture}. Spectroscopic redshift information is used when available, and photometric redshift estimates if not.

\section{Results}

\subsection{Retrospective Classification}
Our model achieves $99.73 \pm 0.26$\% accuracy, $99.68 \pm 0.35$\% precision, $99.78 \pm 0.22$\% recall, and an AUC of 0.9994 on the Ia vs. non-Ia binary classification test dataset, prepared as described in Section 2.1.  \cref{binary-confusion} shows the confusion matrix, created with data from five independent runs of the classifierwith a classification threshold of 0.5.

\begin{figure*}[ht]
  \begin{center}
  \includegraphics[scale=0.28,trim={0 0 0 0}]{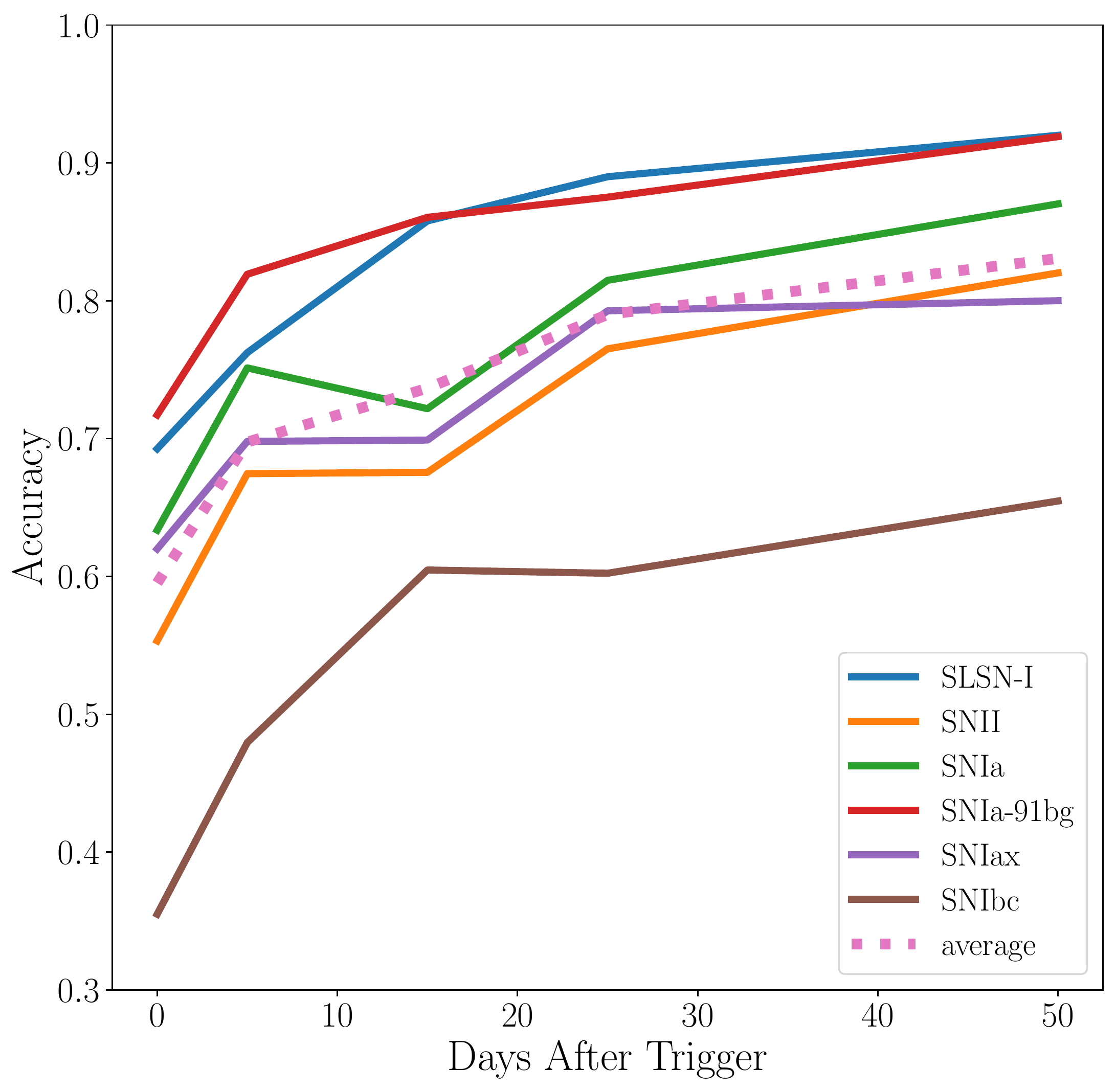}
  \includegraphics[scale=0.28,trim={0 0 0 0}]{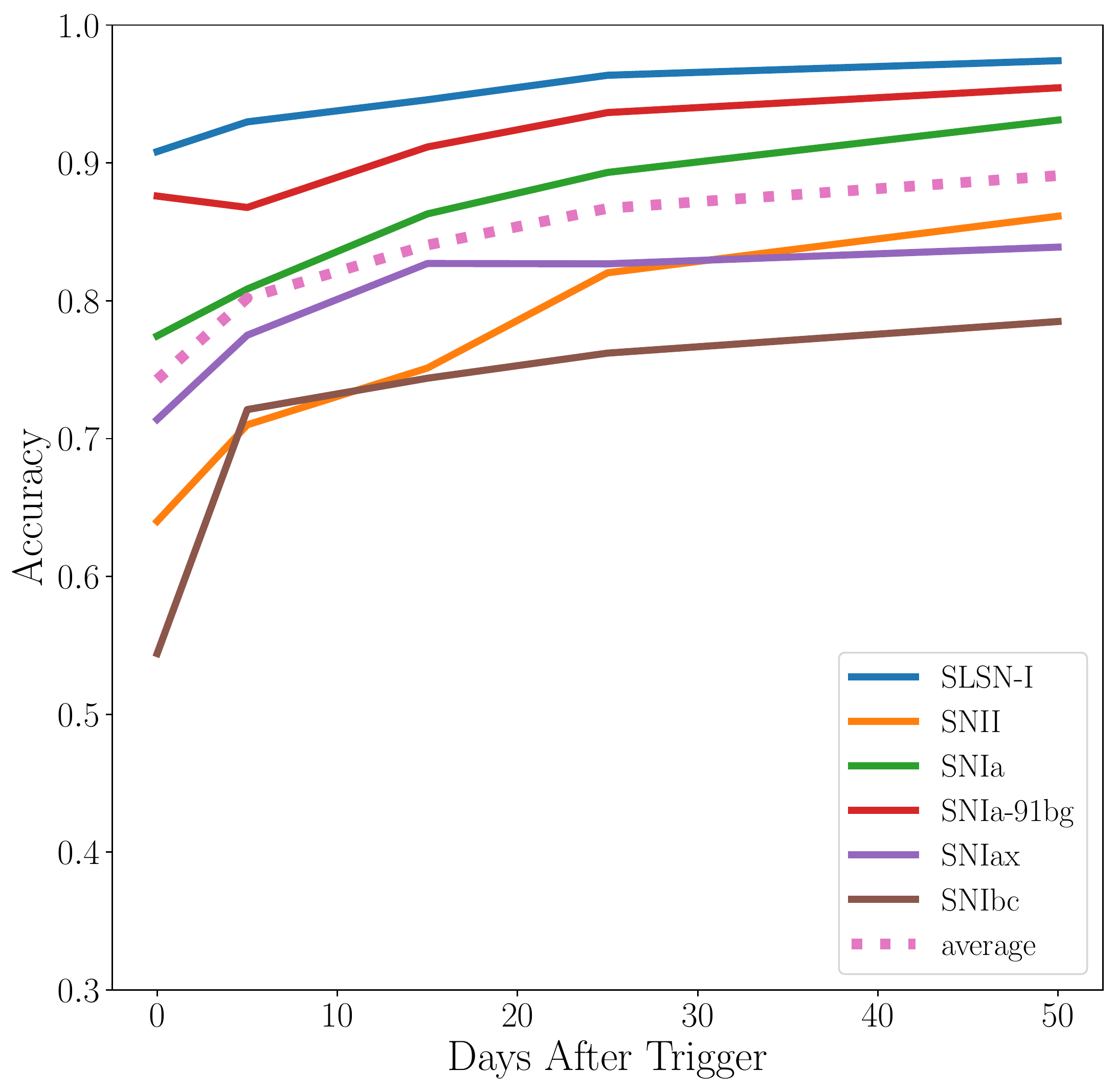}

  \centering
  \caption{Accuracy/efficiency over time for each supernova type without redshift (top) and with redshift (bottom) for the $t_{\mathrm{trigger}}+N$ test datasets.}
  \label{accs}
  \end{center}
\end{figure*}
\begin{figure}
    \includegraphics[scale=0.25]{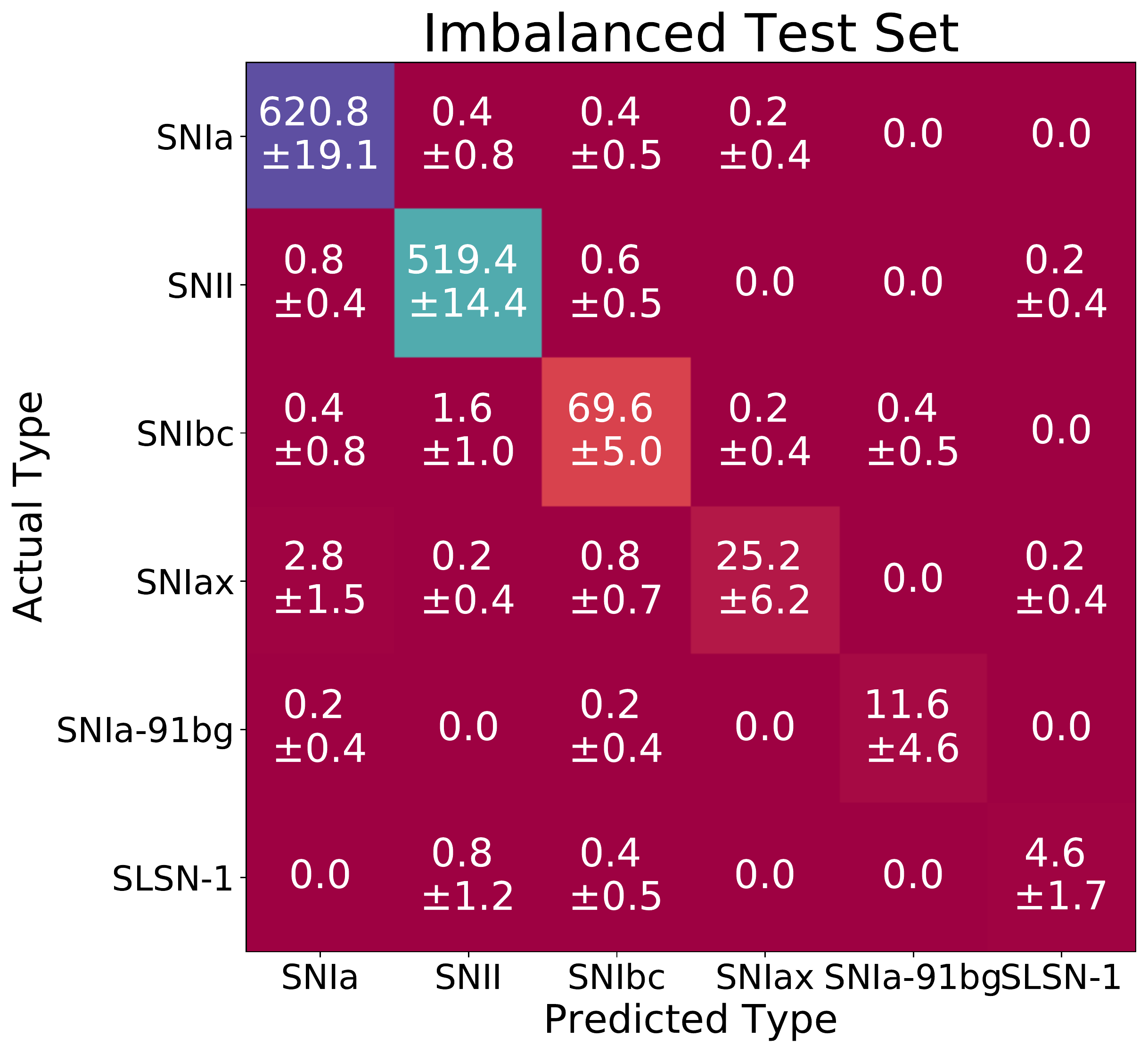}
    \centering
    \caption{Confusion matrix showing average and standard deviation over five runs for SN typing on an imbalanced (realistic) dataset. The absolute numbers of examples are shown here as opposed to percentages.}
    \label{imbalanced-confusion}
\end{figure}

To simulate a more realistic classification scenario, our model was trained and evaluated on a dataset based on observed abundances of each SN class and no class balancing was performed. The model still performs very well under these circumstances, as shown by the confusion matrix in \cref{imbalanced-confusion}.

\subsection{Real-Time Typing}

\cref{accs} shows the accuracy evolution over time for each supernova type in the test sets. The average classification accuracy across all 6 types starts at $\sim 60\%$ without redshift information on the night of trigger, and 75\% with redshift. Redshift unequivocally improves classification performance, especially at early times when there is little photometric data to learn from. The largest improvement in accuracy occurred between 0 and 5 days after trigger for all datasets, likely due to the inclusion of the epoch of peak brightness.

% The confusion matrices for $t_{\mathrm{trigger}}+5$ test set with and without redshift information are shown in Figure~\ref{cm}. This figure shows that the incorporation of redshift information primarily prevents confusion between SLSN-I and SNIbc, decreasing the confusion from 14-16\% to 2-3\%. This is likely because SLSN-I are extremely luminous and thus can be observed at higher redshifts than other SN types, but may be difficult to distinguish from SNIbc with flux-normalized lightcurve data alone.

\subsection{Comparison to Existing Work}
SuperNNova \cite{supernnova} can classify SNe Ia vs. non-Ia with 99.55\% accuracy training on a dataset of $\sim 900,000$ SNe with spectroscopic redshift information. However, our model achieves similar accuracy with just $\sim$ 12,000 SNe and no redshift information. A closer comparison is a 92.4\% accuracy achieved by SuperNNova trained with $\sim$ 50,000 examples and no redshift information, which is noticeably worse than the results presented in this work. 

SuperRAENN \cite{superraenn} is a semi-supervised model trained on real SN candidates, as opposed to the simulated observations used in this work. The most comparable result is shown in \cref{imbalanced-confusion} where the relative abundances of each SN type are based on real observations. The purity values achieved by our model are significantly higher for all the types shared by both works: 0.92 on SLSNe compared with 0.81, 0.993 on SNIa compared with 0.96, and 0.967 on SNIbc compared with 0.33.

RAPID \cite{rapid} can differentiate between 12 transient types, including 7 supernova types, in real time. RAPID was tested on $g$ and $r$ band simulated data, as opposed to the $ugrizY$ bands available for simulations used in this work. Comparing our \cref{accs}, SCONE improves specifically upon RAPID’s SNIbc and SNII classification accuracy. From Figure 7 of M19, 12\% of SNIbc are correctly classified 2 days after detection, compared to SCONE’s 54\% accuracy at the date of trigger. 2 days after detection, SNII is classified at 7\% accuracy by RAPID compared to 64\% accuracy at the date of trigger by SCONE.

\section{Conclusions and Future Work}
We have presented a model addressing one of the most pressing problems in supernova science today: the lack of sufficient spectroscopic resources. Our model not only creates the opportunity for precision cosmology without spectroscopy by ensuring a pure dataset of Type Ia SNe using photometry alone, but also has demonstrated applicability to spectroscopic targeting with its real-time classification performance. The model is crucially able to perform these tasks with only $10^4$ trainable parameters, making it lightweight and quick to train. 

In future work, we would like to expand our model to classify other transient and variable phenomena. We are also interested in understanding and improving the out-of-distribution performance of the model, as data from simulatations often differs from real observations. Finally, since we rarely have accurate spectroscopic redshift information, we would like to evaluate the effect of incorrect redshift estimates on classification performance and its eventual impact on cosmological parameter estimates.


\bibliography{example_paper}
\bibliographystyle{icml2022}
\end{document}